\documentstyle[preprint,aps]{revtex}
\begin{document}

\title{Resonances in 
positronium-rubidium and positronium-cesium scattering} 

\author{Sadhan K. Adhikari}
\address{Instituto de F\'{\i}sica Te\'orica, 
Universidade Estadual Paulista, 
01.405-900 S\~ao Paulo, S\~ao Paulo, Brazil\\}

\date{\today}
\maketitle

\begin{abstract}

Scattering of ortho positronium (Ps)  by cesium and rubidium atoms  has been
investigated employing a three-Ps-state coupled-channel model with
Ps(1s,2s,2p) states using a  time-reversal-symmetric regularized
electron-exchange model potential.  
We find a narrow S-wave singlet resonance at 5.057 eV of width 0.003 eV in
the
Ps-Rb system and at 5.067 eV of width 0.003 eV in the Ps-Cs system.
Singlet P-wave
resonances in both systems are found at 5.3 eV of width 0.4 eV. Singlet 
D-wave structures are  found  at 5.4 eV  in both systems.
The pronounced P- and D-wave resonances in these systems lead to easily detectable 
local minima in the
low-energy elastic cross sections. 
We also report results for elastic and Ps-excitation cross sections for Ps
scattering by Rb and Cs.

{\bf PACS Number(s):  34.10.+x, 36.10.Dr}

\end{abstract}


\newpage

Recent measurements of ortho positronium
(Ps) scattering cross section by H$_2$, N$_2$, He, Ne, Ar, C$_4$H$_{10}$, and
C$_5$H$_{12}$ \cite{r1,r2,r3,r4,r5,r6,r7} have 
initiated new theoretical
investigations 
\cite{r8,r9,r10,r11,r12,r13,r13x} in this subject. 
We suggested \cite{r14,r15,r16} a regularized, symmetric,  nonlocal
electron-exchange
model potential and used it  
in the
 study of Ps scattering by H \cite{r17}, He \cite{r15,r16,r18},
Ne
\cite{r18}, Ar \cite{r18}, H$_2$ \cite{r19},  Li \cite{r20}, Na \cite{pus} and K
\cite{pus}. The
low-energy
cross sections obtained in these studies 
are in
agreement with experimental measurements   
 for He \cite{r3,r4} , Ne \cite{r7}, Ar \cite{r7} and H$_2$ \cite{r3,r4}.
These
investigations also produced 
correct results for resonance and binding energies for the S
wave electronic
singlet state of Ps-H \cite{r9,r10,r11,r12,r17} and 
Ps-Li \cite{r20} systems in addition to experimental pick-off quenching
rate in
Ps-He \cite{r16} scattering. These studies also predicted 
low-energy resonances  in lower
partial waves
of the Ps-Na and Ps-K systems \cite{pus}.

In this Letter we apply  the above symmetric model-exchange   potential to
study
Ps-Rb and
Ps-Cs scattering using the three-Ps-state coupled-channel method.  The
interaction in the singlet channel is attractive in nature as in the
corresponding channel of the Ps-H system \cite{r17} and  we find
resonances in this channel at low energies in S, P and D waves of
both systems near the Ps(2)  excitation threshold. We also report
angle-integrated elastic and Ps-excitation cross sections for both
systems.

The  resonances  
in electron-atom \cite{r21,r22,r23} and positron-atom
\cite{r24}
systems, and in other atomic systems  in general, are  of great
interest. 
Several resonances in the  
electron-hydrogen
system  have been found in  the close-coupling calculation and later
reconfirmed in the variational calculation \cite{r25,r26,r27}. Resonances
have also
been found in the close-coupling calculation of electron scattering by
Li, Na
and K \cite{r28,r29}.  These resonances provide the
necessary testing ground for a theoretical formulation and  can
eventually be detected experimentally. A proper  dynamical description of
scattering  in a theoretical formulation is necessary
for the appearance of these resonances. The ability of the present  
exchange potential to reproduce the resonances in diverse Ps-atom
systems \cite{r17,r20,pus} assures its realistic nature.

For target-elastic 
scattering 
we solve the following 
Lippmann-Schwinger scattering integral
equation in momentum space \cite{r15,r17} \begin{eqnarray} f^\pm_{\nu
',\nu} (
{\bf k',k})&=& {\cal B}^\pm _{\nu ',\nu}({\bf k ',k})  \nonumber
\\ &-&\sum_{\nu ''} \int \frac{ {\makebox{d}{\bf k''}} } {2\pi^2}
\frac { {\cal B}^ \pm _ {\nu ', \nu''} ({\bf k ',k''}) f^ \pm
_{ \nu'' ,\nu} ({\bf k'',k}) } {{k}^2_{\nu ''}/4-k''^2/4+
\makebox{i}0}, \label{4} \end{eqnarray} where the singlet (+) and triplet
($-$)
``Born"
amplitudes, ${\cal B}^\pm$, are given by $
 {\cal B}^\pm_{\nu ',\nu}({\bf k',k}) =
 g^D_{\nu ',\nu}({\bf
k',k})\pm g^ E_{\nu ',\nu}({\bf k',k}),$
   where $g^D$ and $g^E$ represent the direct and exchange Born amplitudes
and the $f^ \pm$ are the singlet and triplet scattering amplitudes,
respectively. The quantum states are labeled by the indices $\nu$,
 referring to the Ps atom. The
variables ${\bf k}$, ${\bf k'}$, ${\bf k''}$ etc denote the appropriate
momentum states of Ps; ${\bf k}_{\nu ''}$ is the on-shell relative
momentum of Ps in the channel $\nu ''$.  We use
units $\hbar = m = 1$ where $m$ is the electron mass.

To avoid
the complication
of calculating exchange potential with a many-electron wave function,
we consider a frozen-core one-valence-electron
approximation for the targets Rb and Cs. Such wave
functions have been successfully used for scattering of 
alkali metal atoms in other contexts and also for positronium scattering
by Li \cite{r13,r13x}.
The Rb(5s) and Cs(6s) frozen-core hydrogen-atom-like wave functions are
taken  as 
\begin{eqnarray}\label{na}
\phi_{\mbox{Rb}}({\bf r})= \frac{1}{300\sqrt 5\sqrt{4\pi\bar
a_0^3}}(120-240\rho+120\rho^2-20\rho^3+\rho^4)e^{-\rho/2},
\end{eqnarray}
\begin{eqnarray}\label{k}
\phi_{\mbox{Cs}}({\bf r})= \frac{1}{2160\sqrt 6\sqrt{4\pi\bar
a_0^3}}(720-1800\rho+1200\rho^2-300\rho^3+30\rho^4-\rho^5)e^{-\rho/2},
\end{eqnarray}
 where $\rho=2r\alpha $ with 
$\alpha=1/(n\bar a_0)$. Here    $n=5$ for Rb and = 6 for Cs and $\bar 
a_0=(2n^2E_i)^{-1}a_0$ with $E_i$ the ionization energy of the target in
atomic unit and $a_0$ the Bohr radius of H. Here we use
the following experimental values for  ionization energies for Rb and
Cs, respectively: 4.176 eV and 3.893 eV \cite{r30}.

In this coupled-channel calculation we  employ the above  frozen-core
approximation for
the target wave function. In addition we shall neglect the excited states of the
target and employ only the lowest Ps(2) excitations of the Ps.  Although, such an
approximation is not entirely realistic, specially when these excitations are
energetically open, 
it makes this  complicated scattering
process mathematically tractable. Moreover, previous experience with Ps-Li, Ps-Na,
and Ps-K systems in similar three-Ps-state model has revealed interesting physics in
producing resonances and correct binding energies \cite{r20,pus}. The reproduction of
correct
binding energies  assures  of physically plausible low-energy cross sections. 
Hence we believe that the present study of Ps-Rb and Ps-Cs  scattering should lead to
physically reasonable results. However, it would be interesting to repeat this
investigation
in the future including the excited states of the target, as has been by Ray in the
Ps-Li system \cite{r13x}, as well as of Ps,  and compare the results for
low-energy scattering with the present investigation.

The  direct Born  amplitude of Ps scattering is given by \cite{r15}
\begin{eqnarray}\label{1x}
g^D_{\nu',\nu} ({\bf k_f,k_i})&= &
\frac{4}{Q^2}
\int \phi^*({\bf r})\left[ 1-\exp ( \makebox{i} {\bf Q.
r})\right]\phi({\bf r})
\makebox{d}{\bf r}\nonumber \\ &\times&
\int \chi^*_{\nu '}({\bf  t })2\mbox{i} \sin ( {\bf Q}.{\bf  t
}/
2)\chi_{\nu}({\bf  t } ) \makebox{d}{\bf  t },
\end{eqnarray}
where $\phi({\bf r})$ is the target wave function and $\chi({\bf t})$ is
the Ps wave function.
The (parameter-free) exchange amplitude corresponding to the  model potential   
is given by \cite{r17} \begin{eqnarray}\label{1}
g^E_{\nu',\nu} ({\bf k_f,k_i})&= &
\frac{4(-1)^{l+l'}}{D}
\int \phi^*({\bf r})\exp ( \makebox{i} {\bf Q. r})\phi({\bf
r})
\makebox{d}{\bf r}\nonumber \\ &\times&
\int \chi^*_{\nu '}({\bf  t })\exp ( \makebox{i}{\bf Q}.{\bf  t }/
2)\chi_{\nu}({\bf  t } ) \makebox{d}{\bf  t },
\end{eqnarray}with
$D=(k_i^2+k_f^2)/8+\alpha^2+(\beta_\nu^2+
\beta_{\nu'}^2)/2$
where $l$  and $l'$ are  the angular momenta of the initial and final Ps
states. The initial and
final Ps momenta are ${\bf k_i}$  and ${\bf k_f}$, ${\bf Q = k_i -k_f}$,
  and $\beta_\nu^2$
and $\beta_{\nu '}^2$ are the binding energies of the initial and final
states of   Ps in atomic unit,
respectively.  
The exchange potential
for Ps scattering is considered
\cite{r14}  to be a
generalization 
of the Ochkur-Rudge exchange potential for electron scattering
\cite{r33,r34}.
It is possible to introduce an adjustable parameter in the above exchange potential
to fit the scattering result to an accurately known low-energy observable, such as
the Ps-atom
scattering length or  binding energy \cite{r18}. However, in Ps-Rb and Ps-Cs systems
there is no
such observable and we use the parameter-free exchange amplitude (\ref{1}) in this
investigation.

After a partial-wave projection, the system of coupled equations (\ref{4}) 
is solved by the method of matrix inversion. 
 Forty Gauss-Legendre quadrature points are
used in the discretization of
each momentum-space integral.  The calculation is performed with the
exact Ps wave functions and frozen-core orbitals (\ref{na}) and (\ref{k})
for Rb(5s) and Cs(6s) ground
state. We consider  Ps-Rb and Ps-Cs  scattering using the
three-Ps-state model 
that includes the 
 following states: 
Ps(1s)Rb(5s), Ps(2s)Rb(5s), Ps(2p)Rb(5s), and 
Ps(1s)Cs(6s), Ps(2s)Cs(6s),
Ps(2p)Cs(6s), for Rb and Cs, respectively.

The Ps-Rb and Ps-Cs systems have an effective attractive interaction in
the  singlet channel as in the Ps-H \cite{r17} and Ps-Li \cite{r20}
systems. The targets of these systems have one active valence electron
outside a
closed shell.  In the coupled-channel calculation we find resonances in both systems
in the singlet state.
 No resonances appear in the triplet state.
For the resonances to appear, the inclusion of the excited states of Ps is
fundamental in a coupled-channel calculation.  The static-exchange model
with both the target and Ps in the ground state does not lead to these
resonances. A detailed investigation of these resonances in coupled-channel model
of Ps-H \cite{r9,r10,r11,r17} and Ps-Li \cite{r20} systems in the singlet
state
has appeared in the literature. 

Here to study  the resonances, first we calculate the S-, P- and D-wave
elastic phase shifts and cross sections in the singlet channel of the
Ps-Rb and Ps-Cs systems
using the 3-Ps-state model.  The phase shifts are calculated in the usual fashion
\cite{bran} from the partial-wave scattering amplitude or from the partial-wave $S$
matrix. The energy and width of resonance are obtained by fitting the corresponding
partial-wave cross section to the well-known
Breit-Wigner formula
\begin{equation}
\sigma (E) \sim \frac{\pi}{k^2} \frac {\Gamma ^2 }{(E-E_R)^2+ \Gamma^2/4},
\end{equation}  
where $\sigma (E)$ is the cross section at  energy $E=6.8 k^2$ eV,
$E_R$ is the resonance energy and $\Gamma$  the width.
 The S-wave phase shifts are shown in Fig.
1. The Ps-Rb system has a  resonance at 5.057 eV of width 0.003 eV. The
resonance in the Ps-Cs system appears at 5.067 eV and also has a width
0.003
eV.  The phase shift curves in Fig.  1
clearly show the resonances where the phase shifts jump by $\pi$.

In Fig. 2 we show the P-wave Ps-Rb and Ps-Cs phase shifts in the singlet
state. Both systems possess resonances at 5.2 eV of width of 0.3 eV. The
P-wave singlet elastic cross section at low energies shown in the off-set
clearly exhibits these resonances.  In Fig. 3 we plot the D-wave singlet
elastic cross section for Ps-Rb  and Ps-Cs systems at low energies. There
is
a structure in both systems at 5.4 eV which is more diffuse than in S and
P waves. 

Next we calculate the different partial cross sections of Ps-Rb and Ps-Cs
scattering. The convergence of the cross sections with respect to partial
waves is slower in
this case than in the case of Ps-H scattering. At a incident Ps energy of
50 eV, 40 partial waves were used to achieve convergence of the
partial-wave scheme. In Figs.  4 and 5 we plot different angle-integrated
cross sections of Ps-Rb and Ps-Cs scattering, respectively. Specifically,
we plot the elastic, Ps(2s) and Ps(2p) excitation cross sections using the
three-Ps-state model. For comparison we also plot the elastic cross
section obtained with the static-exchange model. The elastic cross section
is  large at low energies in both systems. In the off-set of Figs. 
4 and 5 we plot the corresponding low-energy elastic cross sections.  The
effect of the inclusion of highly polarizable Ps(2) states in the coupling
scheme could be considerable, specially at low energies.

From Figs. 4 and 5 we see that the P and D-wave 
resonances of large widths 
predicted in this Letter have lead to local minima in the elastic 
cross
section below the Ps excitation threshold at 5.1 eV.  Hence these resonances can be
easily detected 
experimentally after an analysis of the elastic scattering cross section at low
energies. This makes these resonances of great experimental interest. Similar minima 
appear in the cross section for electron scattering by alkali-metal atoms due to
appearance of resonances in these systems \cite{r18,r29}. 
However, the
S-wave
resonances in the Ps-Rb and Ps-Cs systems  are narrow and may not be easily noted
experimentally from a simple analysis of the cross sections.

To summarize, we have performed a three-Ps-state coupled-channel
calculation of Ps-Rb and Ps-Cs scattering at low energies using a
regularized symmetric nonlocal electron-exchange model potential 
\cite{r14,r15} successfully used \cite{r15,r17,r18,r19,r20} previously in
different Ps scattering problems. We present the results of
angle-integrated partial cross
sections at different Ps energies.  We also present results for
singlet phase
shifts and partial-wave cross sections at low energies to study
the resonances in these systems. We find resonances in S, P and D waves
near the Ps(2) excitation threshold.
In this Letter we have used a three-Ps-state model. Similar resonances have
been found in the coupled-channel model of electron-H
\cite{r25,r26,r27},
electron-Na, electron-K \cite{r28,r29}, positron-hydrogen \cite{r24}, Ps-H
\cite{r10,r11,r17} and Ps-Li \cite{r20} systems. In most cases, a more
complete
calculation and (in some cases) experiments have reconfirmed these
resonances. In view of this we do not believe that the appearance of these 
resonances in the present three-state calculation to be so peculiar as to
have no general validity. However, the resonance energies might change
slightly after a more complete calculation as in electron-H,
positron-H and Ps-H systems and it would be
intersting to study the present resonances using more complete theoretical
models in the future in addition to compare the present results with
future experiments.

The work is supported in part by the Conselho Nacional de Desenvolvimento -
Cient\'\i fico e Tecnol\'ogico,  Funda\c c\~ao de Amparo
\`a Pesquisa do Estado de S\~ao Paulo,  and Finan\-ciadora de Estu\-dos e
Projetos of Brazil.

{\bf Figure Caption:}

1. Singlet S-wave elastic phase shift at different Ps energies for Ps-Rb
(dashed
line) and Ps-Cs (full line) scattering.

2. Singlet P-wave elastic phase shift at different Ps energies for Ps-Rb
(dashed
line) and Ps-Cs (full line) scattering. The corresponding P-wave singlet
cross section is shown in the off-set.

3. Singlet D-wave elastic cross section  at different Ps energies for
Ps-Rb
(dashed line) and Ps-Cs (full line) scattering. 

4. Partial cross sections for Ps-Rb scattering at different Ps energies:
three-Ps-state elastic (full line), three-Ps-state Ps(2s) (dashed-dotted
line), three-Ps-state Ps(2p) (short-dashed line), static-exchange elastic
(long-dashed line). Three-Ps-state (full line) and static-exchange
(long-dashed line) elastic  results at low energies are also shown in the
off-set.

5. Same as in Fig. 4 for Ps-Cs scattering.

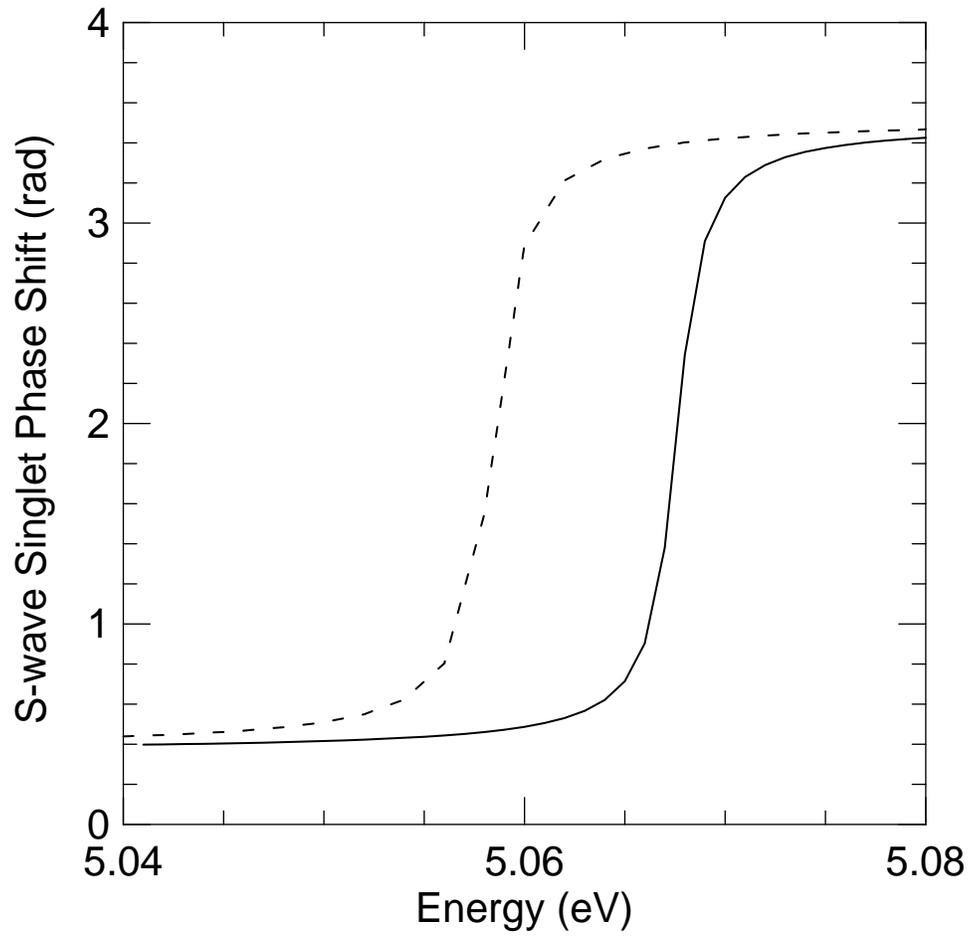
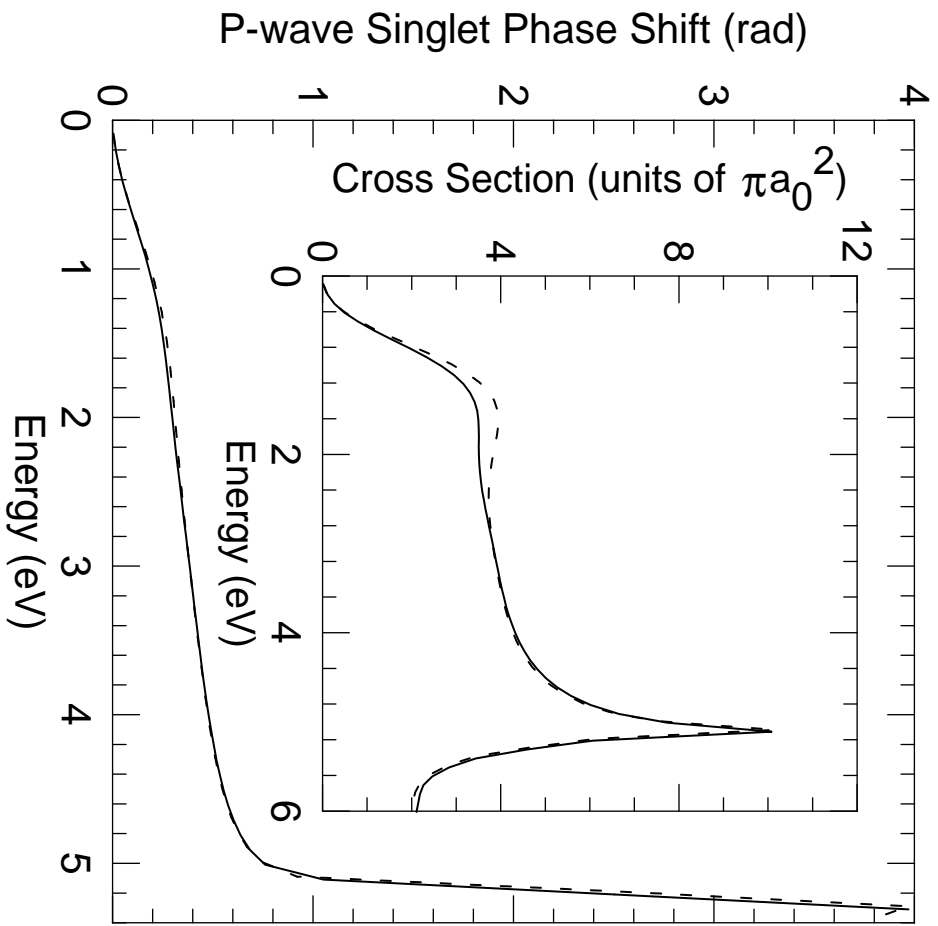
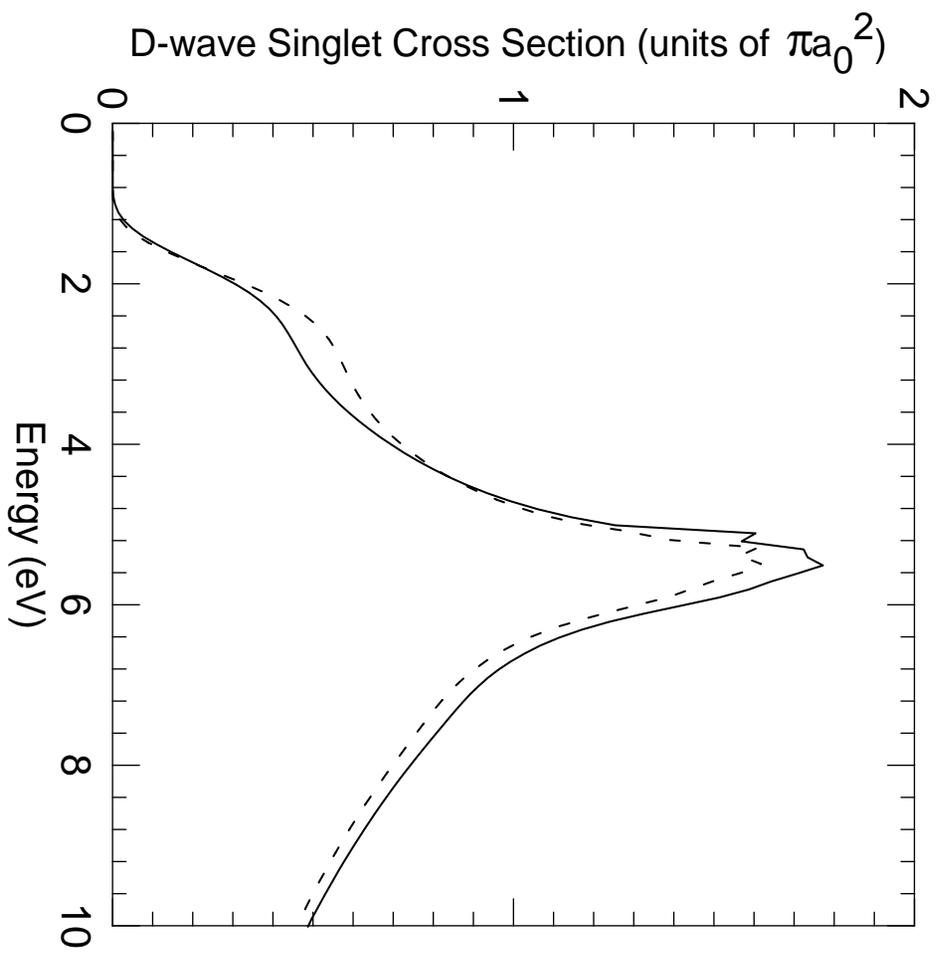
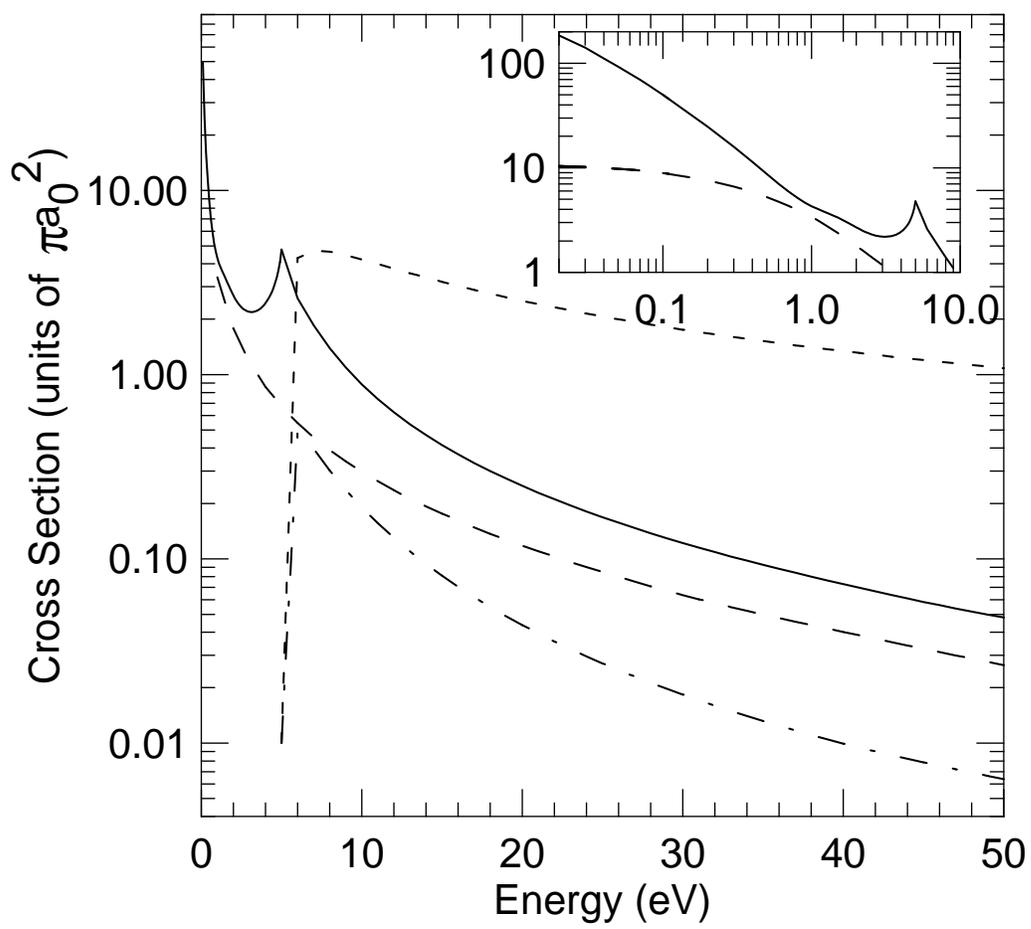
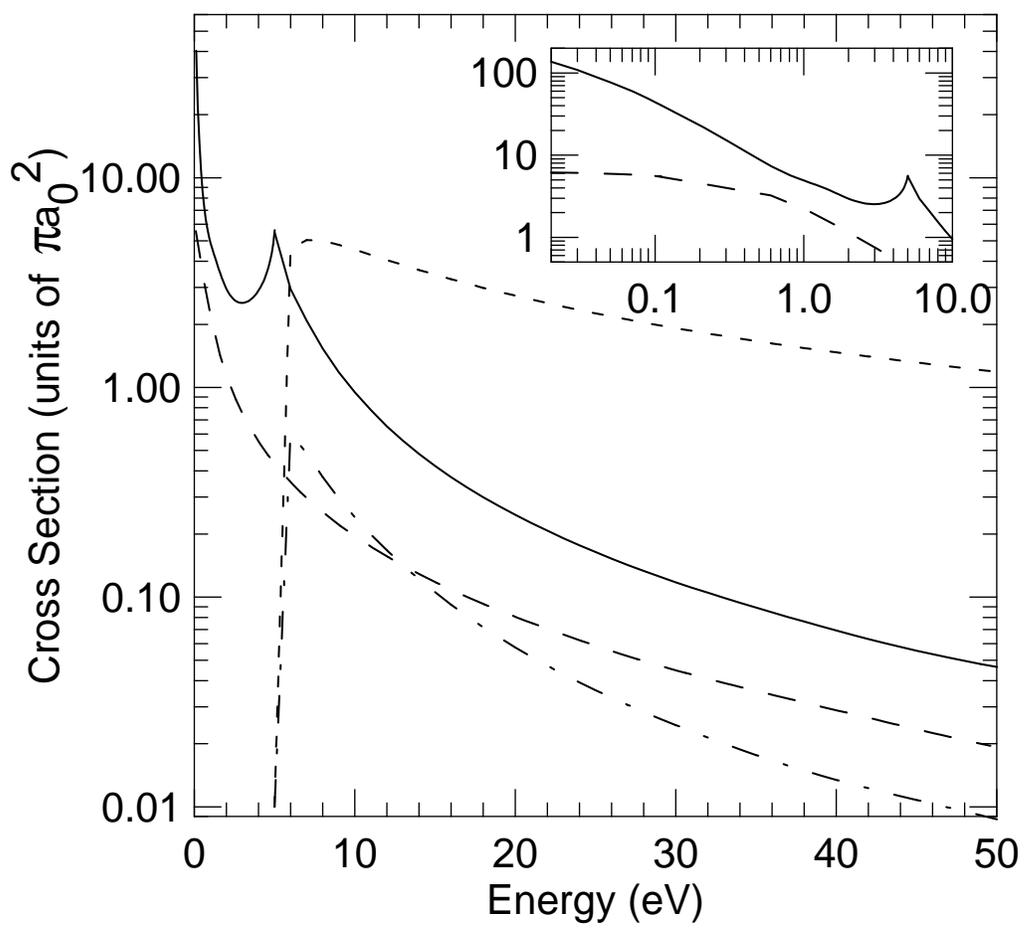
\end{document}